\documentclass{article}
\usepackage{spconf,amsmath,amssymb,graphicx,float,hyperref}
\usepackage{booktabs}
\usepackage[table]{xcolor}
\usepackage{multirow}
\usepackage{tabularx}

\title{Long-Prompt Zero-Shot TTS: Benchmarking Prosody Reproduction Beyond Voice Cloning}

\name{Dekun Chen$^{1}$, Renjie Xu$^{1}$, Yingda Shen$^{1}$, Qinke Ni$^{1}$, Zhizheng Wu$^{1,2}$\sthanks{Corresponding author.}}
\address{$^{1}$The Chinese University of Hong Kong, Shenzhen
\qquad $^{2}$Amphion Technology Co., Ltd.}

\begin{document}
\ninept
\maketitle

\begin{abstract}

Zero-shot text-to-speech (TTS) typically conditions on a short speech prompt, which conveys timbre but provides only limited evidence of a speaker's prosodic patterns.
We formulate \emph{Long-Prompt Zero-Shot TTS}, a task that uses long speech prompts at inference time to reproduce speaker-specific prosody without target-speaker fine-tuning.
We introduce LP-TTS-Bench, a bilingual benchmark of 37 speakers with 63.95 prompt hours and 2,728 held-out test utterances.
Beyond intelligibility and timbre similarity, LP-TTS-Bench evaluates timing, prominence, intonation, and model-based prosody metrics, enabling multidimensional evaluation of prosody reproduction.
As a reference system, we develop LPTTS, whose hierarchical Long-Prompt Encoder compresses long speech prompts into a fixed number of continuous prosody vectors.
Experiments show improvements in several prosodic dimensions over a short-prompt baseline, with overall gains as prompt duration increases up to approximately 10 minutes and saturation at longer durations.

\end{abstract}

\begin{keywords}
Text-to-speech, long-prompt TTS, zero-shot synthesis, speaker-specific prosody
\end{keywords}

\section{Introduction}
\label{sec:introduction}

Recent zero-shot text-to-speech (TTS) systems can reproduce a target speaker's voice from only a few seconds of speech, without speaker-specific parameter updates~\cite{wang2023valle,du2024cosyvoice2,hu2026qwen3tts}.
Because timbre is relatively stable within a speaker, such short prompts often capture voice identity well.
A speaker's manner of speaking, however, is more complex: speech rate, pauses, emphasis, and intonation vary with communicative context and personal habits so that a short prompt reveals only a local realization rather than recurring patterns across contexts.
Prior work has also treated speaker identity and prosody as distinct attributes~\cite{lux2023exact}.
The improved prosody similarity observed from 3 to 10 seconds of prompt~\cite{shen2024naturalspeech2} further motivates using longer prompts to estimate speaker-specific prosodic behavior.

When additional target-speaker recordings are available, speaker adaptation can incorporate them into model parameters~\cite{chen2021adaspeech}, but requires extra training and management of configuration.
Multi-utterance prompting provides an alternative and has been explored in Mega-TTS 2 and PerTTS~\cite{jiang2024megatts2,li2026pertts}, although their implementations are not publicly available.
Building on this direction, we formulate \emph{Long-Prompt Zero-Shot TTS}: synthesizing speech from minutes to tens of minutes of prompt speech without updating target-speaker parameters.
Our central question is whether and how additional prompt material improves the reproduction of speaker-specific prosody as prompt duration increases.

\begin{figure}[t]
    \centering
    \includegraphics[width=\columnwidth]{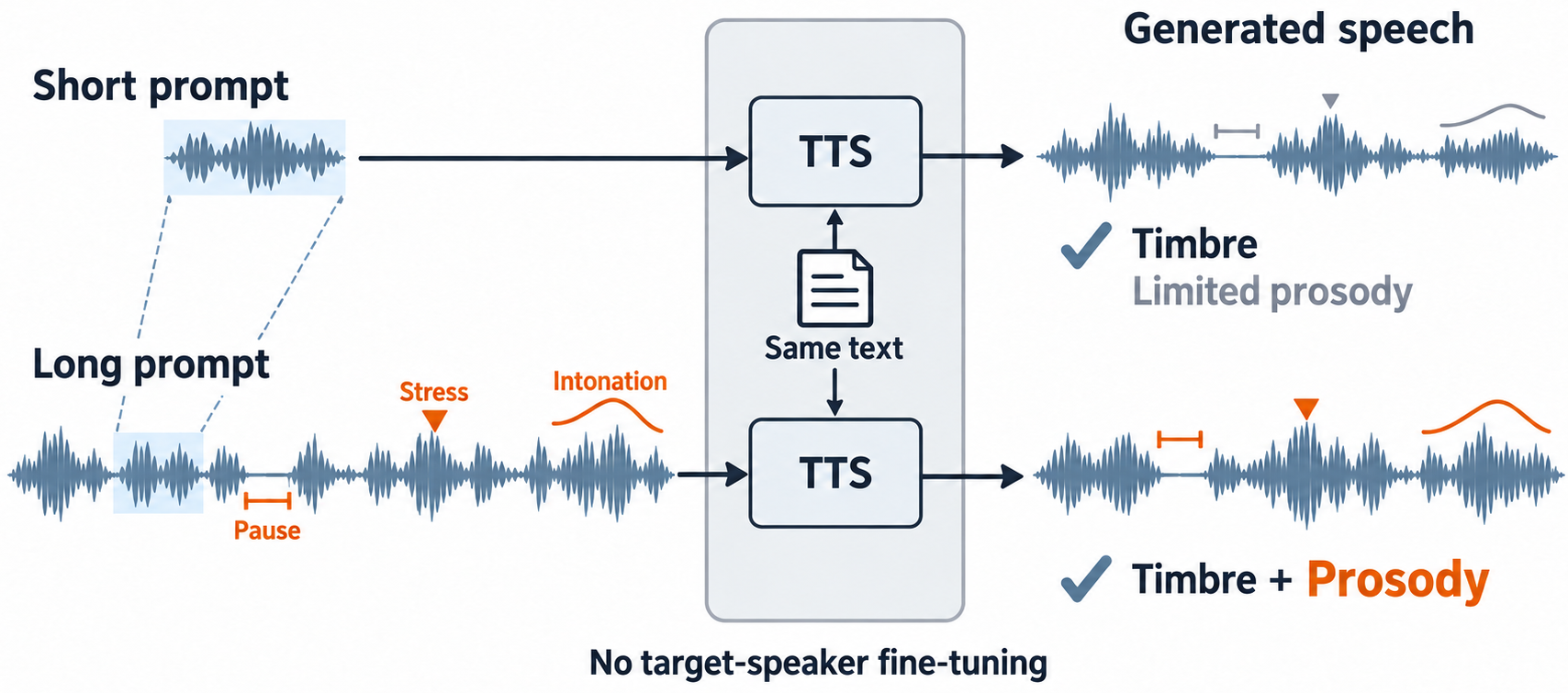}
    \caption{Illustration of \emph{Long-Prompt Zero-Shot TTS}. Compared with a short
    prompt, a long prompt provides richer evidence of speaker-specific prosody,
    enabling the model to reproduce both timbre and prosody without
    target-speaker fine-tuning.}
    \label{fig:motivation}
\end{figure}

Existing TTS evaluation has begun to move beyond intelligibility and speaker similarity.
ProsodyEval, for example, benchmarks prosody diversity using human ratings and an objective metric~\cite{yang2026prosodydiversity}.
Our focus instead is speaker-specific prosody reproduction against matched ground-truth utterances.
We therefore introduce LP-TTS-Bench, a Chinese--English benchmark of 37 speakers with a multidimensional prosodic protocol.

To provide a reference system for this task and benchmark, we develop LPTTS with a hierarchical Long-Prompt Encoder built on Qwen3TTS~\cite{hu2026qwen3tts}.
It compresses variable-length speech prompts into a fixed number of global tokens and target-text-conditioned local tokens.
Training combines speech generation with a same-speaker contrastive ranking objective and then applies on-policy distillation~\cite{agarwal2024gkd}.
Experiments show improvements in several prosodic dimensions, with overall gains up to approximately 10 minutes and saturation at longer prompt durations.

Our contributions are threefold: we (1) formulate Long-Prompt Zero-Shot TTS as a new task, (2) introduce LP-TTS-Bench and its multidimensional evaluation protocol, and (3) provide LPTTS as an initial reference system, together with analyses of prompt duration and text-speech prosody relevance.
LP-TTS-Bench, evaluation scripts, and audio samples will be released upon acceptance.

\section{LP-TTS-Bench}
\label{sec:benchmark}

\begin{figure*}[t]
    \centering
    \includegraphics[width=0.9\textwidth]{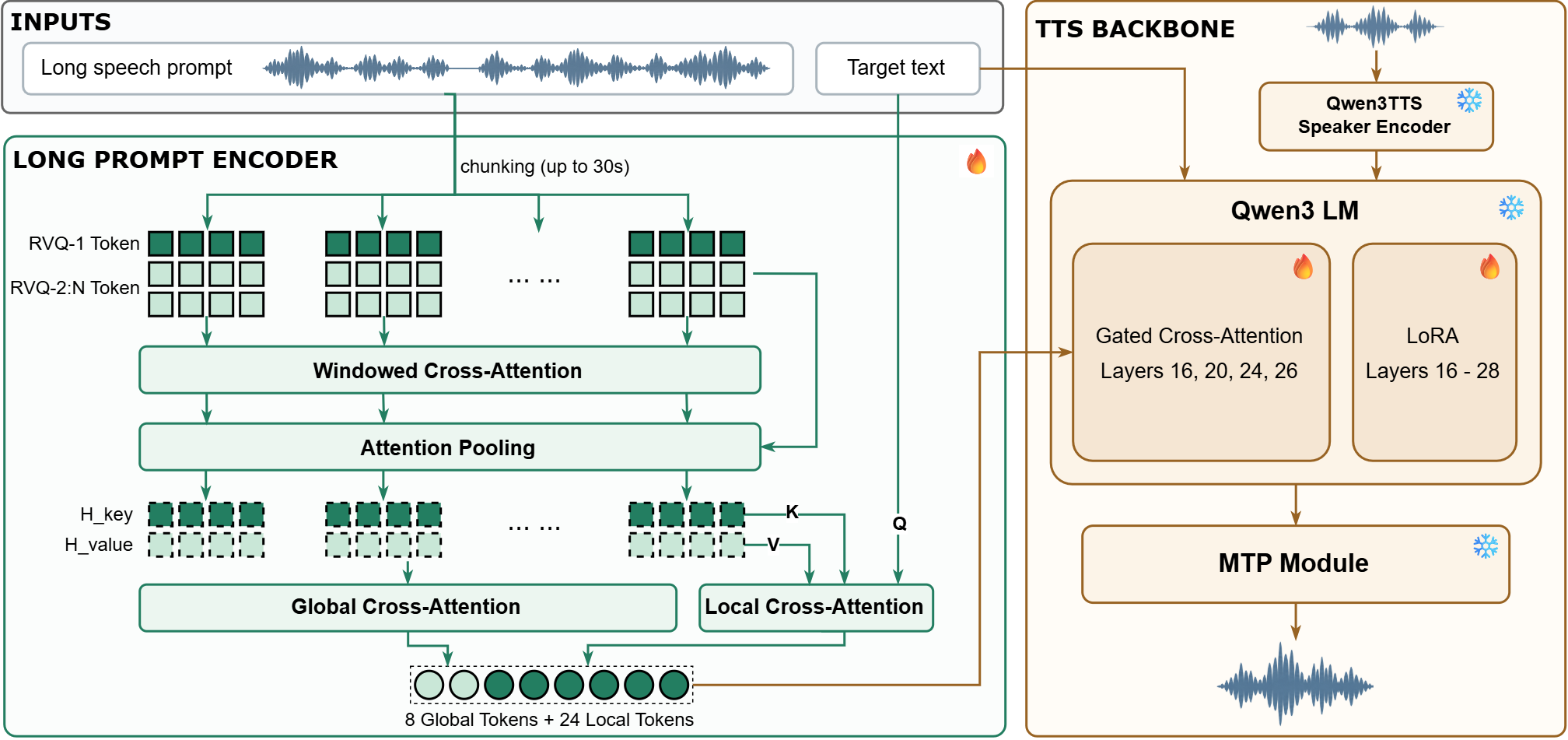}
    \caption{Overview of the LPTTS system.
    The Long-Prompt Encoder extracts global and target-conditioned
    prosodic representations from long speech prompt and injects them
    into Qwen3TTS by gated cross-attention.
    Timbre conditioning is provided by the native speaker encoder.}
    \label{fig:architecture}
\end{figure*}

\subsection{Task Definition and Data}
\label{ssec:task-data}

Long-Prompt Zero-Shot TTS aims to reproduce speaker-specific prosody from extended speech prompts at inference time.
Given target text $x$ and a prompt corpus $P_s$ from speaker $s$, the model synthesizes $\hat{y}=f_{\theta}(x,P_s)$ without target-speaker parameter updates.
We study how increasing the amount of prompt speech affects prosody reproduction across utterances.

LP-TTS-Bench contains 17 Chinese speakers and 20 English speakers, with 63.95 prompt hours, 3,559 validation utterances, and 2,728 test utterances in total.
Each speaker has 57--80 held-out test utterances longer than five seconds.
The speaker corpus is used either to condition long-prompt systems or to adapt speaker-fine-tuning baselines, while the validation set supports training configuration for the latter.
Each test item pairs a target text with its ground-truth recording, enabling controlled comparisons of prompting, fine-tuning, and prompt duration on a fixed test set.

\begin{table*}[t]
\centering
\caption{Main results on LP-TTS-Bench.
WER/CER is reported in percent.
Arrows indicate the preferred direction.
The checkmark indicates target-speaker fine-tuning.
Bold and underlining denote the best and second-best results within each language, respectively; tied values share the same formatting.}
\label{tab:main_results}
\small
\setlength{\tabcolsep}{1pt}
\renewcommand{\arraystretch}{1.0}
\begin{tabularx}{\textwidth}{
    @{} l >{\centering\arraybackslash}p{0.085\textwidth}
    *{2}{>{\centering\arraybackslash}X}
    *{2}{>{\hsize=.85\hsize\linewidth=\hsize\centering\arraybackslash}X}
    *{3}{>{\centering\arraybackslash}X}
    *{2}{>{\hsize=1.15\hsize\linewidth=\hsize\centering\arraybackslash}X}
    >{\centering\arraybackslash}X @{}
}
\toprule
\multirow[c]{2}{*}[-0.3ex]{Model}
& \multicolumn{1}{c}{\multirow[c]{2}{*}[-0.3ex]{\shortstack{Target Spk.\\SFT?}}}
& \multicolumn{2}{c}{Conventional}
  & \multicolumn{2}{c}{Timing}
  & \multicolumn{1}{c}{Prominence}
  & \multicolumn{2}{c}{Intonation}
  & \multicolumn{2}{c}{Prosody model}
  & \multicolumn{1}{c}{Subjective} \\
\cmidrule(lr){3-4}
\cmidrule(lr){5-6}
\cmidrule(lr){7-7}
\cmidrule(lr){8-9}
\cmidrule(lr){10-11}
\cmidrule(lr){12-12}
& & \makebox[0pt][c]{\footnotesize WER/CER $\downarrow$}
& \makebox[0pt][c]{\footnotesize SIM-O $\uparrow$}
& \makebox[0pt][c]{\footnotesize SRD $\downarrow$}
& \makebox[0pt][c]{\footnotesize PD $\downarrow$}
& \makebox[0pt][c]{\footnotesize PCC $\uparrow$}
& \makebox[0pt][c]{\footnotesize IC $\uparrow$}
& \makebox[0pt][c]{\footnotesize PRD $\downarrow$}
& \makebox[0pt][c]{\footnotesize MPM-SIM $\uparrow$}
& \makebox[0pt][c]{\footnotesize AutoPCP $\uparrow$}
& \makebox[0pt][c]{\footnotesize P-MOS $\uparrow$} \\
\midrule
\rowcolor{gray!10}
\multicolumn{12}{c}{\textit{Chinese}} \\
Qwen3TTS-Base & $\times$
& \textbf{0.73} & 0.56
& 0.155 & 0.074 & 0.389
& 0.437 & 5.017 & 0.721 & 3.128 & 2.941 \\
Qwen3TTS-SFT & $\checkmark$
& 1.78 & \underline{0.57}
& \underline{0.139} & \textbf{0.065} & \underline{0.395}
& \underline{0.463} & \underline{4.764} & \underline{0.733} & \textbf{3.216} & \underline{3.471} \\
LPTTS-Base & $\times$
& 1.29 & \textbf{0.58}
& 0.155 & 0.069 & 0.387
& 0.437 & 5.013 & 0.728 & \underline{3.181} & 3.235 \\
LPTTS & $\times$
& \underline{1.03} & \underline{0.57}
& \textbf{0.137} & \underline{0.067} & \textbf{0.410}
& \textbf{0.474} & \textbf{4.532} & \textbf{0.734} & 3.148 & \textbf{3.529} \\
\midrule
\rowcolor{gray!10}
\multicolumn{12}{c}{\textit{English}} \\
Qwen3TTS-Base & $\times$
& \textbf{1.32} & 0.77
& 0.124 & 0.076 & \underline{0.516}
& 0.303 & 3.668 & 0.659 & 3.368 & 3.133 \\
Qwen3TTS-SFT & $\checkmark$
& 1.93 & \underline{0.78}
& \textbf{0.111} & 0.069 & \textbf{0.527}
& \underline{0.318} & \underline{3.345} & \underline{0.677} & 3.443 & \underline{3.267} \\
LPTTS-Base & $\times$
& 1.95 & \textbf{0.79}
& 0.126 & \underline{0.068} & 0.506
& 0.298 & 3.570 & 0.668 & \textbf{3.477} & 3.200 \\
LPTTS & $\times$
& \underline{1.64} & \underline{0.78}
& \underline{0.112} & \textbf{0.061} & 0.514
& \textbf{0.335} & \textbf{3.279} & \textbf{0.681} & \underline{3.452} & \textbf{3.533} \\
\bottomrule
\end{tabularx}
\end{table*}

\subsection{Evaluation Protocol}
\label{ssec:evaluation}

We evaluate nine objective metrics covering intelligibility, speaker similarity, timing, prominence, intonation, and model-based prosody metrics.
Each synthesized utterance is compared with the matched ground-truth recording of the same target text.

\smallskip
\noindent\textbf{Intelligibility and speaker similarity.}
We use Qwen3-ASR~\cite{shi2026qwen3asr} to compute WER for English and CER for Chinese.
SIM-O measures the cosine similarity between ERes2Net speaker embeddings~\cite{chen2023eres2net} extracted from synthesized and ground-truth speech.

\smallskip
\noindent\textbf{Timing.}
Motivated by error-based evaluation of speech-rate estimates~\cite{farrus2024speechrate}, \emph{Speech Rate Distance (SRD)} measures the absolute log-ratio of speaking rates:

$$
r=\frac{N}{t_{\mathrm{last}}^{\mathrm{end}}-t_{\mathrm{first}}^{\mathrm{start}}},
\quad
\mathrm{SRD}=\left|\log\frac{r_{\mathrm{syn}}}{r_{\mathrm{gt}}}\right|,
$$

where $N$ is the number of text units, and the duration includes internal pauses but excludes leading and trailing silence.

Following reference-based analysis of pause placement and duration in synthetic speech~\cite{chan2024prosodic}, \emph{Pause Distance (PD)} compares pause strength at corresponding unit boundaries:

$$
v_i=\log\left(1+\frac{\max(0,g_i-\tau)}{\tau}\right),
\;
\mathrm{PD}=\frac{1}{N-1}\sum_i |v_i^{\mathrm{syn}}-v_i^{\mathrm{gt}}|,
$$

where \(g_i\) is the nonnegative gap between adjacent units and \(\tau=0.15\) s.
This penalizes missing, misplaced, and mismatched-duration pauses.

\smallskip
\noindent\textbf{Prominence.}
Following acoustic prominence models that combine $F_0$, energy, and duration~\cite{suni2017prosody}, \emph{Prominence Contour Correlation (PCC)} computes the Spearman correlation between unit-level scores that equally combine energy, local pitch movement or peak, and relative log-duration after clause-level median/MAD normalization.

\smallskip
\noindent\textbf{Intonation.}
Following objective TTS evaluation using aligned $F_0$-contour correlation~\cite{liu2023prosody}, \emph{Intonation Correlation (IC)} compares smoothed semitone contours sampled at corresponding within-unit positions to reduce duration-induced misalignment.
After clause-level centering and amplitude normalization, it averages the correlations of the contours and their first-order differences:

$$
\mathrm{IC}=\frac{\operatorname{corr}(q^{\mathrm{syn}},q^{\mathrm{gt}})
+\operatorname{corr}(\Delta q^{\mathrm{syn}},\Delta q^{\mathrm{gt}})}{2}.
$$

Following robust percentile-based pitch-range analysis~\cite{ryant2016style}, \emph{Pitch Range Distance (PRD)} measures the mean absolute difference between the clause-level voiced-pitch ranges of the synthesized and ground-truth utterances, where each range is the 90th--10th percentile span in semitones.

\smallskip
\noindent\textbf{Model-based prosody metrics.}
\emph{MPM-SIM} computes cosine similarity between utterance representations obtained by mean and max pooling Masked Prosody Model features~\cite{wallbridge2025mpm}.
\emph{AutoPCP}~\cite{seamless2023} predicts perceptual prosodic similarity on a 1--4 scale.

Qwen3-ForcedAligner~\cite{shi2026qwen3asr} aligns each synthesized and ground-truth pair to the shared target text for timing, prominence, and intonation metrics.
Chinese characters and English words serve as text units, while punctuation defines clause boundaries.
Pitch is expressed in semitones relative to the speaker's median voiced \(F_0\), and relative duration and normalized energy reduce sensitivity to speaking rate and recording gain.
For paired comparisons, the five alignment-based metrics are computed on the common subset of test utterances for which all evaluated systems achieve utterance-level WER/CER below 0.1, reducing interference from content errors while preserving a matched evaluation set.

\section{LPTTS System}
\label{sec:method}

Fig.~\ref{fig:architecture} gives an overview of LPTTS, our reference system for LP-TTS-Bench, including the hierarchical Long-Prompt Encoder and its gated integration with the Qwen3TTS backbone.

\subsection{Long-Prompt Encoder Design}
\label{ssec:long-prompt-encoder}

Built on Qwen3TTS-Base~\cite{hu2026qwen3tts}, the Long-Prompt Encoder receives features from all 16 RVQ codebooks of the native speech tokenizer.
It divides the long speech prompt into chunks of up to 30 seconds, compresses each chunk, and aggregates the resulting representations into 32 fixed-length conditioning vectors, independent of prompt duration.
Of these vectors, 8 global tokens summarize overall prosodic patterns, while 24 local tokens are retrieved using target-text queries and semantic keys from the first RVQ codebook.
Our design combines global style representations~\cite{wang2018styletokens}, hierarchical prosody modeling~\cite{huang2022generspeech}, and text-conditioned prompt retrieval~\cite{xue2024retrieval}, while using full-codebook features to retain acoustic information.
Section~\ref{ssec:text-prosody-ablation} provides empirical support for this design by showing that textually similar utterances tend to exhibit greater prosodic similarity.

Motivated by prior work that separates timbre from prosody or style~\cite{jiang2024megatts2,lee2026fctts}, we use the native Qwen3TTS speaker encoder for timbre, leaving the Long-Prompt Encoder to model prosody.
We concatenate the global and local tokens into $Z$ and inject them into Qwen3 LM using gated cross-attention inspired by Flamingo~\cite{alayrac2022flamingo}:
$$
H^{(\ell)} \leftarrow H^{(\ell)}
+\gamma_{\ell}\operatorname{CrossAttn}_{\ell}
\bigl(H^{(\ell)},Z,Z\bigr),
$$
where $H^{(\ell)}$ denotes the Qwen3 LM hidden states at layer $\ell$ and $\gamma_{\ell}$ is a learnable gate.
The hidden states supply the queries, while the 32 conditioning vectors supply the keys and values.
We insert these modules at layers 16, 20, 24, and 26, with gates initialized to 0.01.

\subsection{Training Objectives}
\label{ssec:training-objectives}

\smallskip
\noindent\textbf{Encoder training.}
We first train on a mixture of short and long speech prompts: 30\% of batches use prompts shorter than 30 seconds, and 70\% use prompts from one to 30 minutes.
Alongside the cross-entropy (CE) loss, we apply a same-speaker contrastive ranking objective to 40\% of long-prompt batches.
For two long prompts $P^{+}$ and $P^{-}$ from the same speaker, target utterances are sampled from the recording associated with $P^{+}$.
The target utterance and its surrounding segment are excluded from \(P^{+}\), ensuring that the ranking objective cannot exploit direct acoustic or textual overlap between the prompt and the target.
Let $\mathcal{L}^{+}$ and $\mathcal{L}^{-}$ denote the CE losses obtained by conditioning on the respective prompts while keeping the target text and speaker embedding fixed.
The ranking loss and total training objective are
\[
\mathcal{L}_{\mathrm{rank}}=\max\bigl(0,0.05+\mathcal{L}^{+}-\mathcal{L}^{-}\bigr),
\quad
\mathcal{L}=\mathcal{L}_{\mathrm{CE}}+0.2\mathcal{L}_{\mathrm{rank}}.
\]
Because \(P^{+}\) and \(P^{-}\) share speaker identity while the target utterances are expected to be prosodically closer to \(P^{+}\), this objective holds timbre conditioning constant and encourages the encoder to distinguish prompt-specific prosodic information.
During this stage, all parameters of the Long-Prompt Encoder and gated cross-attention (GCA) modules are trained, while Qwen3 LM is adapted using LoRA in 16 layers; all remaining parameters are frozen.

\smallskip
\noindent\textbf{On-policy distillation.}
During on-policy distillation (OPD)~\cite{agarwal2024gkd}, a Qwen3TTS teacher fine-tuned on approximately 20 hours from one expressive speaker supervises a prompt-conditioned student initialized from the encoder-training checkpoint.
Long prompts are constructed by randomly concatenating utterances from the same speaker.
On self-generated trajectories, the student minimizes the next-token Jensen--Shannon divergence from the teacher, while only the Long-Prompt Encoder and its injection modules are updated; all TTS-backbone parameters, including LoRA, remain frozen.

\section{Experiments and Analysis}
\label{sec:experiments}

\subsection{Experimental Setup}
\label{ssec:experimental-setup}

We use Qwen3TTS-Base~\cite{hu2026qwen3tts} as the backbone.
Encoder training uses approximately 50K hours of in-the-wild, speaker-paired Chinese and English recordings with prompt durations from 30 seconds to 30 minutes.
OPD then follows Section~\ref{ssec:training-objectives}, using an expressive game-voice speaker who is absent from LP-TTS-Bench.
Benchmark speakers are disjoint from the 50K-hour encoder-training data and the OPD speaker.
Prompt, validation, and test samples are mutually exclusive, and test utterances are never included in the long prompts.

The baselines are Qwen3TTS-Base with a 10-second prompt and Qwen3TTS-SFT.
Encoder training produces LPTTS-Base and applying OPD to this checkpoint produces LPTTS.
Both LPTTS variants use 10-minute prompts and require no target-speaker training.
Qwen3TTS-SFT instead comprises 37 checkpoints, each independently adapted via full fine-tuning for five epochs on the corresponding LP-TTS-Bench speaker's complete training corpus, with checkpoint selection based on the speaker-specific validation set and evaluation restricted to that speaker.
All systems synthesize the same test texts and use the metrics in Section~\ref{ssec:evaluation}.
For P-MOS, 100 test items balanced across languages and speakers were evaluated in a randomized, system-blind test, with each output rated for prosodic similarity to the matched ground-truth recording on a 1--5 scale by three listeners proficient in the corresponding language.

\begin{figure}[t]
    \centering
    \includegraphics[width=\columnwidth]{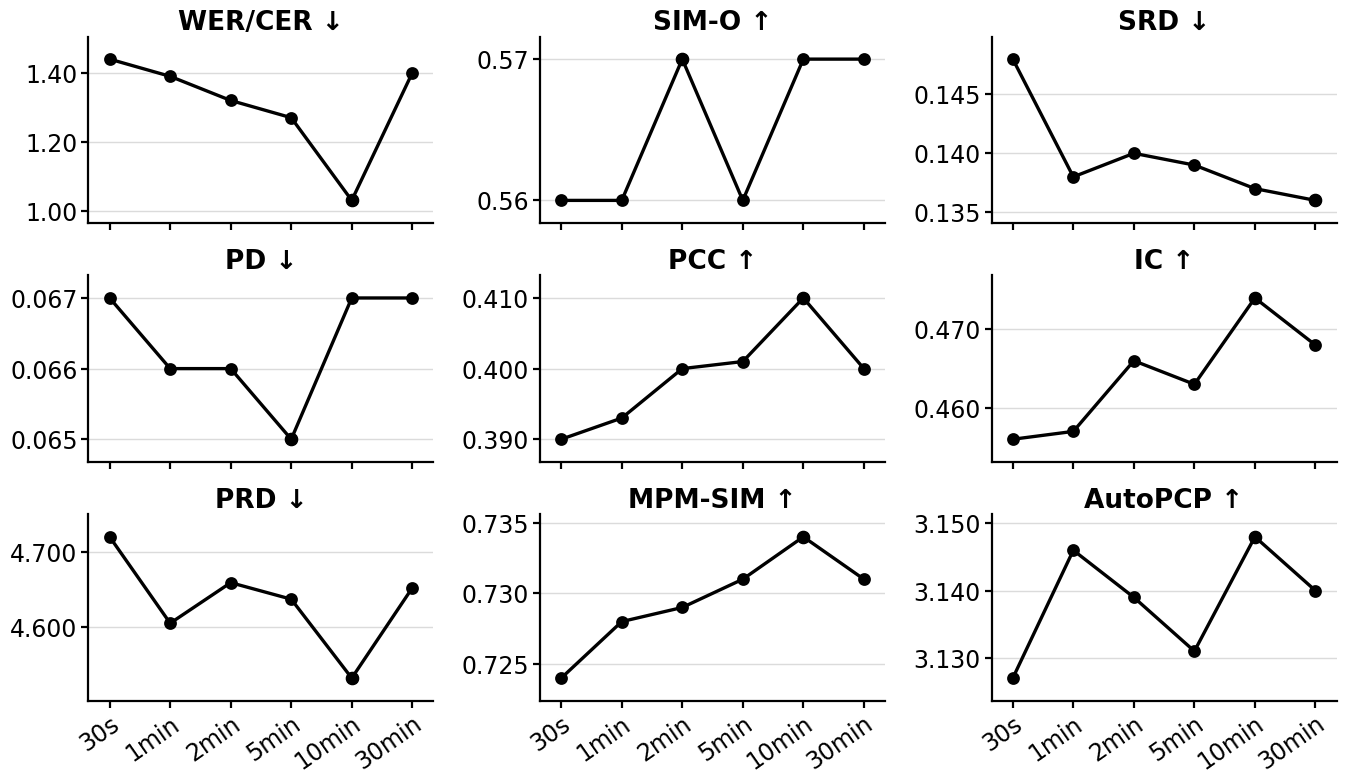}
    \caption{Effect of prompt length on intelligibility,
    speaker similarity, and prosody reproduction for Chinese.
    English results show similar trends and are omitted due to space limitations.}
    \label{fig:prompt_length}
\end{figure}

\subsection{Main Results}
\label{ssec:main-results}

Table~\ref{tab:main_results} compares the four systems across the objective metrics and P-MOS.
Compared with Qwen3TTS-Base, LPTTS-Base achieves broadly similar performance on the explicit prosodic metrics despite using a substantially longer prompt.
PD and the model-based prosody metrics improve, whereas SRD, PCC, and IC show limited or inconsistent changes.
This suggests that encoder training primarily enables the model to process and integrate long-prompt representations, while its ability to extract useful prosodic information remains limited.

Qwen3TTS-SFT improves every prosody-related metric over Qwen3TTS-Base, demonstrating the benefit of direct target-speaker adaptation.
However, WER/CER deteriorates from 0.73 to 1.78 in Chinese and from 1.32 to 1.93 in English.
This trade-off suggests that fitting target-speaker characteristics can reduce pretrained intelligibility and also requires separate training for each speaker.

After OPD, LPTTS improves over LPTTS-Base across all five explicit prosody metrics and MPM-SIM in both languages.
It also outperforms Qwen3TTS-Base on most prosody metrics.
Because the OPD speaker is absent from LP-TTS-Bench, these gains support generalization beyond the training speaker.
The results suggest that distillation strengthens the encoder's ability to extract prosodic information from long prompts rather than merely accommodating their length.
Although WER/CER remains above Qwen3TTS-Base, OPD reduces recognition errors relative to LPTTS-Base.
Finally, P-MOS is broadly consistent with the objective trends: LPTTS ranks highest, LPTTS-Base and Qwen3TTS-SFT form a comparable middle tier, and Qwen3TTS-Base ranks lowest.
Overall, without target-speaker fine-tuning, LPTTS achieves better numerical scores than Qwen3TTS-SFT on six of eight prosody metrics in each language, together with lower WER/CER and comparable speaker similarity, offering a single-model alternative to per-speaker adaptation.

\subsection{Effect of Prompt Length}
\label{ssec:prompt-length}

To isolate the effect of prompt duration from model and training differences, we evaluate the same LPTTS checkpoint while varying only the amount of prompt speech available at inference time.
For each speaker, shorter prompts are nested prefixes of the corresponding longer prompts.

Fig.~\ref{fig:prompt_length} summarizes how the evaluation metrics change as prompt duration increases from 30 seconds to 30 minutes.
Across prompt lengths, the metrics related to speech rate, prominence, intonation, and MPM-SIM improve overall, typically reaching their best values around 10 minutes rather than increasing monotonically.
Meanwhile, SIM-O remains approximately stable, suggesting that the prosodic gains are not accompanied by substantial changes in timbre similarity.
This pattern is consistent with the encoder's intended focus on prosodic information.
Overall, increasing prompt duration provides additional prosodic evidence, although the gains are non-monotonic and tend to saturate beyond 10 minutes.

\begin{table}[t]
\centering
\caption{Relationship between textual and prosodic similarity.
Negative $\Delta$ favors textually similar prompts.}
\label{tab:text_prosody}
\small
\setlength{\tabcolsep}{3pt}
\renewcommand{\arraystretch}{0.9}
\begin{tabular*}{0.9\columnwidth}{
    @{\extracolsep{\fill}} l c c @{}
}
\toprule
Feature & $\Delta$ [95\% CI] & Closer Rate \\
\midrule
Pitch
& $-0.054$ [$-0.061$, $-0.040$]
& 60.2\% \\
Energy
& $-0.047$ [$-0.060$, $-0.036$]
& 56.4\% \\
Timing/pause
& $-0.065$ [$-0.082$, $-0.063$]
& 59.9\% \\
\midrule
Overall
& $-0.059$ [$-0.067$, $-0.053$]
& 62.6\% \\
\bottomrule
\end{tabular*}
\end{table}

\subsection{Text--Prosody Relationship Analysis}
\label{ssec:text-prosody-ablation}

To motivate target-text-conditioned retrieval, we conduct this analysis on the 37-speaker LP-TTS-Bench and, for each anchor, select three textually similar utterances and ten random utterances while matching text length and audio duration.
We then measure pitch, energy, and timing/pause distances to the anchor.
Table~\ref{tab:text_prosody} reports \(\Delta=d_{\mathrm{similar}}-d_{\mathrm{random}}\) and the proportion of comparisons in which the textually similar prompts are closer.
Every feature group has a negative difference with a 95\% confidence interval excluding zero ($p<0.001$ for all comparisons), and the overall Closer Rate is 62.6\%.
These results show that textually similar utterances tend to exhibit more similar prosody, supporting target-text-conditioned retrieval as a meaningful way to extract prosodic evidence from long prompts that is relevant to the target linguistic context.

\section{Conclusion}
\label{sec:conclusion}

We formulate Long-Prompt Zero-Shot TTS and introduce LP-TTS-Bench, a bilingual 37-speaker benchmark for multidimensional evaluation of speaker-specific prosody reproduction.
Experiments with LPTTS show that longer prompts, particularly up to 10 minutes, improve most prosody metrics over short-prompt conditioning, while further increases exhibit diminishing or non-monotonic gains.
Together, these findings demonstrate both the value of long-prompt conditioning and the need to evaluate speaker-specific prosody beyond conventional intelligibility and speaker-similarity measures.
LP-TTS-Bench thus provides a testbed for future research on long-context, speaker-specific prosody modeling.

\newpage

\section{Compliance with Ethical Standards}
The listening evaluation involved human participants.
All participants provided informed consent, and no personally identifiable information was collected.
The study was conducted in accordance with the applicable institutional guidelines.

\section{Acknowledgments}
This work is partially supported by the National Natural Science Foundation of China under Grant 62376237,
and the Program for Guangdong Introducing Innovative and Enterpreneurial Teams (Grant No. 2023ZT10X044).

\bibliographystyle{IEEEbib}
\bibliography{strings,refs}

\end{document}